\documentclass[aps,amssymb,amsfonts,amsmath,twocolumn]{revtex4}
\usepackage{times}
\usepackage{revsymb}
\usepackage{bm}
\begin{document}
\small
\normalsize
\protect\newtheorem{principle}{Principle}
\protect\newtheorem{theo}[principle]{Theorem}
\protect\newtheorem{prop}[principle]{Proposition}
\protect\newtheorem{lem}[principle]{Lemma}
\protect\newtheorem{co}[principle]{Corollary}
\newtheorem{cri}[principle]{Criterion}
\protect\newtheorem{de}[principle]{Definition}
\protect\newtheorem{con}[principle]{Conjecture}
\newtheorem{ex}[principle]{Example}
\newtheorem{rema}[principle]{Remark}
\newtheorem{rem}[principle]{Remark}
\small
\normalsize
\title{Some
Properties of the Computable Cross Norm Criterion for Separability}
\author{Oliver Rudolph}
\email{rudolph@fisicavolta.unipv.it}
\affiliation{Quantum Optics \& Information Group, Istituto Nazionale per la Fisica
della Materia \& Dipartimento \\ di Fisica ``A.~Volta,''
Universit\`a degli Studi
di Pavia, via Bassi 6, I-27100 Pavia, Italy}
\begin{abstract}
\noindent The computable cross norm (\texttt{CCN}) criterion is a new powerful analytical
and computable separability criterion for bipartite quantum states,
that is also known to
systematically detect bound entanglement. In certain aspects this
criterion complements the well-known Peres positive partial transpose
(\texttt{PPT}) criterion.
In the present paper we study important analytical
properties of the \texttt{CCN} criterion.
We show that in contrast to the \texttt{PPT} criterion
it is not sufficient in dimension 2
$\times$ 2. In higher dimensions we prove theorems connecting
the fidelity of a quantum state with the \texttt{CCN} criterion.
We also analyze the behaviour of the \texttt{CCN} criterion
under local operations
and identify the operations that leave it invariant. It turns out
that the \texttt{CCN} criterion is in general
not invariant under local operations.
\end{abstract}
\maketitle

\section{Introduction}
{Entanglement of composite quantum systems
is a key resource in many applications of quantum
information technology.} However, theoretically entanglement is
not yet fully understood and to decide whether or not a given
state is entangled or useful for quantum information processing
purposes is in general a difficult question.
Therefore the
characterization and classification of entangled states is an
important area of research
that has received much attention in the development
of quantum information theory.
In recent years considerable progress has been made towards developing a
general theory of quantum entanglement.
In particular criteria to decide whether or not a
given quantum state
is entangled are of high theoretical and practical interest.
Historically, Bell type inequalities were the first
operational criteria to distinguish between entangled and separable
states. Due to the importance of entanglement
in quantum information processing there has been a dramatic
increase in our knowledge and understanding of entangled quantum
states. Today, we have much more subtle and effective separability
criteria than provided by Bell inequalities.
Most notably, in Ref.~\onlinecite{Peres96} Peres obtained
a powerful computable necessary
separability criterion, the so-called positive partial transpose (\texttt{PPT})
criterion.
The Peres criterion
stipulates that the partial transpose of any separable quantum
state is again a state.
The Horodecki family formulated
a necessary and sufficient mathematical characterization
of separable states in terms of positive maps \cite{Horodecki96b}.
Subsequently, the study of separability criteria and
their relation to positive maps
attracted a great deal of
attention and several new criteria were formulated \cite{Bruss02}.
By now there exists a sophisticated theory based on so-called
entanglement witnesses \cite{Bruss02,Terhal,Eckert02}.
However, for a long time the \texttt{PPT} criterion remained the
most powerful and versatile operational separability criterion.
It was only relatively
recently that a novel \emph{analytical}
separability criterion not based
on entanglement witnesses or positive maps was derived in
Ref.~\onlinecite{Rudolph02}. The new criterion was
derived within the context of an approach that aims to characterize
entanglement by using norms \cite{Rudolph00}. In Ref.~\cite{Rudolph02}
the new criterion was named
\emph{computable cross norm criterion} for reasons to become clear
below. In the present paper we shall adopt this terminology and
for brevity also use the acronym \emph{\texttt{CCN} criterion}.
The \texttt{CCN} criterion is as easy to compute and
as versatile as the \texttt{PPT}
criterion, but yet independent of it \cite{Rudolph02}.
The new criterion is the first analytical separability criterion
that is known to systematically
detect bound
entanglement as well as genuine multipartite entanglement
\cite{Horodecki02}. The power of the new criterion was
already demonstrated
in Ref.~\onlinecite{Rudolph02} where a number of examples
were discussed.
It was shown there that the
\texttt{CCN} criterion is necessary and sufficient for pure states while
for mixed states the \texttt{CCN} criterion is not
sufficient in dimension $d \geq 3$. For dimension $2 \times 2$ the question
of sufficiency was left open.

Recently a non-analytical but computationally tractable
generalization of the \texttt{PPT} criterion
based on semidefinite programming was
presented in Ref.~\onlinecite{Doherty02}. This powerful method is
also able to detect bound entanglement. It is clear,
however, that the same ideas can also be applied to the \texttt{CCN}
criterion. It is therefore natural to conjecture
that the tests described in Ref.~\onlinecite{Doherty02}
together with the analogue generalization of the \texttt{CCN}
criterion will provide a very powerful hierarchy of numerical separability
tests.

The \texttt{CCN} criterion complements the Peres criterion in several
aspects. The aim of the present paper is to
study and clarify some important analytical
properties of the \texttt{CCN} criterion in detail.
We shall demonstrate three important results.
In Section \ref{s3} we study the \texttt{CCN} criterion
in dimension 2 $\times$ 2. We find that the criterion is in general
not sufficient in
dimension 2 $\times$ 2. We also  prove that for two qubit states with
maximally disordered subsystems the \texttt{CCN} criterion is
necessary and sufficient.
In Section \ref{s4}
we study the \texttt{CCN} criterion in arbitrary dimension and prove
theorems relating upper and lower bounds for the fidelity of
quantum states to the \texttt{CCN} criterion.
Finally in Section \ref{s5} we study the behaviour of the \texttt{CCN} criterion
under local operations.
We show that the \texttt{CCN} criterion is not invariant under
local operations and therefore also not under
\texttt{LQCC} operations (i.e.,
quantum operations that can be implemented locally with classical
communication between the parties). We put forward a generalization of the \texttt{CCN}
criterion that is strictly stronger than the
\texttt{CCN} criterion.
In the course of the present paper we
employ key techniques and methods that we hope
will prove useful also for
further studies and applications of the \texttt{CCN} criterion.

Throughout the paper we adopt the following notation: the set of
bounded operators on ${\mathbb{C}}^d$ (i.e., $d \times d$ matrices)
is denoted by
${\mathtt{T}}({\mathbb{C}}^d)$. The canonical real basis of
${\mathbb{C}}^d$ is denoted by $(\vert i \rangle)_{i=1}^d$ and the
maximally entangled wavefunction with respect to this basis is
denoted by $\vert \Psi_+ \rangle \equiv \frac{1}{\sqrt{d}}
\sum_{i=1}^d \vert i i \rangle$.

\section{The {CCN} Criterion}
A quantum state $\varrho$ on ${\mathbb{C}}^d \otimes
{\mathbb{C}}^d$ is called \emph{separable} (disentangled) if it can
be expressed as a
convex combination of product states \cite{Werner89}, i.e., in the
form
\[ \varrho = \sum_{i=1}^k p_i \varrho_i \otimes \tilde{\varrho}_i. \]
Otherwise $\varrho$ is called \emph{entangled}.

The \texttt{CCN} criterion is a necessary separability
criterion. It can be formulated in different equivalent ways.
A very useful and instructive way is the following procedure.
Consider a quantum state $\varrho$
defined on a tensor product Hilbert space ${\mathbb{C}}^d \otimes
{\mathbb{C}}^d$. We denote the canonical real
basis in ${\mathbb{C}}^d$ by
$(\vert i \rangle)_{i=1}^d$ and expand $\varrho$ in terms of
the operators $E_{ij} \equiv \vert i \rangle \langle j \vert$,
i.e., we write
\begin{equation} \label{e1}
\varrho = \sum_{ijkl} \varrho_{ijkl} E_{ij} \otimes E_{kl}.
\end{equation}
Next, we define an operator ${\mathfrak{A}}(\varrho)$ that acts on
${\mathtt{T}}({\mathbb{C}}^d \otimes {\mathbb{C}}^d)$ by
\begin{equation} \label{e2}
{\mathfrak{A}}(\varrho) \equiv \sum_{ijkl} \varrho_{ijkl} \vert
E_{ij} \rangle \langle E_{kl} \vert.
\end{equation}
Here $\vert E_{ij} \rangle$ denotes the ket vector with respect to
Hilbert-Schmidt inner product $\langle A, B \rangle \equiv
{\mathrm{tr}}(A^\dagger
B)$ in ${\mathtt{T}}({\mathbb{C}}^d)$. We also write
$\Vert A \Vert_2 \equiv \langle A, A \rangle^{1/2}.$ The norm
$\Vert A \Vert_2$ is often called the \emph{Hilbert-Schmidt norm} or
the \emph{Frobenius norm} of $A$ and is equal to the sum of the squares
of the singular values of $A$. The sum of the absolute values of
the singular values of $A$ is called the \emph{trace class norm,} or
simply \emph{trace norm,} and is denoted by $\Vert A \Vert_1$.
\begin{cri} The \texttt{CCN} criterion
asserts that if $\varrho$ is separable, then the trace
class norm of ${\mathfrak{A}}(\varrho)$ is less than or equal to
one. Whenever a quantum state $\varrho$ satisfies
$\Vert {\mathfrak{A}}(\varrho) \Vert_1 > 1$, this signals that $\varrho$
is entangled. \end{cri}
In Ref.~\onlinecite{Rudolph02} it has been shown that the criterion is
independent of the basis of ${\mathbb{C}}^d$ chosen.
In fact, there is the following representation for $\Vert {\mathfrak{A}}(\varrho) \Vert_1$
\begin{equation} \label{tau}
\tau(\varrho) \equiv \Vert {\mathfrak{A}}(\varrho) \Vert_1 = \inf \left\{ \sum_{i}
\Vert x_i \Vert_2 \Vert y_i \Vert_2 : \varrho = \sum_i x_i \otimes
y_i \right\} \end{equation} where the infimum runs over all decompositions of
$\varrho$ into finite sums of simple tensors. It is easy to see
that the norm $\tau$ satisfies the inequality
\[ \tau(\sigma_1 \otimes \sigma_2) \leq \Vert \sigma_1 \Vert_1
\Vert \sigma_2 \Vert_1. \]

\noindent This inequality is called the \emph{subcross property} in the
mathematical literature, which justifies the name \emph{computable
cross norm criterion}. From Equations \ref{e1} and \ref{e2} it is
a straightforward and trivial exercise to determine the matrix
representation for ${\mathfrak{A}}(\varrho)$ in the canonical basis. It turns out that
${\mathfrak{A}}(\varrho)$ is equal to the so-called \emph{Oxenrider-Hill matrix
reordering} of $\varrho$ that was
studied in Ref.~\onlinecite{OxenriderH85}.

We conclude this section by remarking that also
the Peres criterion can be
written in the form of a norm criterion. I.e., the Peres criterion is
equivalent to the following statement: if a state $\varrho$ satisfies
$\Vert \varrho^{{\mathtt{T}}_2} \Vert_1 > 1$, then $\varrho$ is
entangled. Here ${\mathtt{T}}_2$ denotes the partial transpose
with respect to the second subsystem.

\section{\label{s3}The {CCN} criterion for two qubits}
\noindent In Ref.~\onlinecite{Rudolph02} the \texttt{CCN} citerion was computed for several
examples, including Werner states, isotropic and Bell diagonal
states. In dimension 2 $\times$ 2
the \texttt{CCN} criterion turned out to be necessary and
sufficient for all these examples.
It is the purpose of this section to study the \texttt{CCN} criterion in dimension
2 $\times$ 2 in more detail.
It is known that any two qubit state $\varrho$ can be
expressed in terms of Hilbert-Schmidt operators, \begin{widetext}
\begin{equation} \label{e3} \varrho = \frac{1}{4} \left( {\openone} \otimes
{\openone}
+ {\mathbf{r}} \cdot \bm{\sigma}
\otimes {\openone} +
{\openone} \otimes {\mathbf{s}} \cdot \bm{\sigma} +
\sum_{m,n=1}^3 t_{mn} \sigma_n \otimes \sigma_m \right). \end{equation}
\end{widetext}
Here $\openone$ stands for the identity operator, $\{ \sigma_i
\}_{i=1}^3$ are the standard Pauli matrices,  ${\mathbf{r}}, {\mathbf{s}} \in
{\Bbb R}^3$ and ${\mathbf{r}} \cdot \bm{\sigma} =
\sum_{i=1}^3 r_i \sigma_i$. We denote the real matrix formed
by the coefficients $t_{mn}$ by $T(\varrho)$.
The separability and distillability
properties of two qubit states in the Hilbert-Schmidt space
formalism have been discussed in detail in
Ref.~\onlinecite{Horodecki967} and \onlinecite{Horodecki96}.
Here we built on these results to study properties of the \texttt{CCN}
criterion. First we note that $\mathbf{r}$ and ${\mathbf{s}}$
equal the Bloch vectors of the reductions $\varrho_1 \equiv
{\mathrm{tr}}_2 \varrho$ and $\varrho_2 \equiv {\mathrm{tr}}_1
\varrho$ of $\varrho$ respectively. A state with maximally
disordered subsystems thus has ${\mathbf{r}} = {\mathbf{s}} =0$ in
Equation \ref{e3}. We prove that the \texttt{CCN} criterion is
necessary and sufficient for two qubit states with maximally
disordered subsystems.
\begin{prop} \label{p2}
Let $\varrho$ be a two qubit state with maximally disordered
subsystems. Then $\Vert {\mathfrak{A}}(\varrho) \Vert_1 = \frac{1 + \Vert T(\varrho)
\Vert_1}{2}$, i.e.,  $\Vert {\mathfrak{A}}(\varrho) \Vert_1 \leq
1$ if and only if $\varrho$ is separable.
\end{prop}
\emph{Proof}: Since the Hilbert-Schmidt norm $\Vert
\cdot \Vert_2$ is invariant under unitaries,
it is obvious from the variational expression for
$\Vert {\mathfrak{A}}(\varrho) \Vert_1$ given above
that $\Vert {\mathfrak{A}}(\varrho)
\Vert_1$ is invariant under local unitary operations of the form
$U_1 \otimes U_2$ acting on $\varrho$. As shown in
Ref.~\onlinecite{Horodecki96} we can always choose local unitaries $U_1,
U_2$ such that $T(U_1 \otimes U_2 \varrho U_1^\dagger \otimes
U_2^\dagger)$ is diagonal. These two facts imply that without loss
of generality we can assume that $T(\varrho)$ is diagonal. Then
$\varrho$ is of the form $\varrho = \frac{1}{4} \left( {\openone} \otimes {\openone}
+ \sum_{m=1}^3 t_{m} \sigma_m \otimes \sigma_m \right)$.
Correspondingly, we find ${\mathfrak{A}}(\varrho) = \frac{1}{4}
\left\vert
\openone \right\rangle \left\langle
\openone \right\vert + \sum_{m=1}^3
\frac{t_{m}}{4}
\left\vert \sigma_m \right\rangle  \left\langle
\sigma_m^* \right\vert$. Here $^*$  denotes complex conjugation.
Note that $\left( \frac{1}{\sqrt{2}}
\left\vert
\openone \right\rangle, \frac{1}{\sqrt{2}} \left\vert
\bm{\sigma}
\right\rangle \right)$ is an orthonormal basis  with respect to
the Hilbert-Schmidt inner product.
$\Vert T(\varrho) \Vert_1$ is invariant under local
unitary operations acting on $\varrho$.
Thus $\Vert {\mathfrak{A}}(\varrho)
\Vert_1 = \frac{1 + \sum_{m=1}^3 \vert t_m \vert}{2} = \frac{1
+ \Vert T(\varrho)
\Vert_1}{2}$. Clearly if $\varrho$ is separable, then $\Vert
{\mathfrak{A}}(\varrho)
\Vert_1 \leq 1$. If $\varrho$ is not separable, then it
follows from Proposition 4 in Ref.~\onlinecite{Horodecki96} that
$\Vert T \Vert_1 > 1$.
This implies $\Vert {\mathfrak{A}}(\varrho) \Vert_1 > 1$.
Alternatively, the last implication also follows from Theorem 2 in
Ref.~\onlinecite{PittengerR00}.
$\Box$

We now wish to relate the \texttt{CCN} criterion with the fidelity
of two qubit states. The fidelity of a state $\varrho$ is defined as
$f(\varrho) \equiv
\max_{\Psi} \langle \Psi \vert \varrho \vert \Psi \rangle$ where
the maximum is over all maximally entangled pure states $\Psi$. The
fidelity is an important quantity that is often employed as a measure
of the efficiency of quantum communication protocols. We
have
\begin{prop} \label{p3}
For any two qubit state $\varrho$ we have $f(\varrho) \leq \frac{\Vert
{\mathfrak{A}}(\varrho) \Vert_1}{2}$.
\end{prop}
The proof of proposition \ref{p3} can be found in Appendix
\ref{a2}.
\begin{prop} \label{p4} Let $\varrho$ be
an entangled two qubit state with
maximally disordered subsystems. Then $\Vert {\mathfrak{A}}(\varrho)
\Vert_1 = 2 f(\varrho)$. \end{prop}
\emph{Proof}: Let $\varrho$ be an entangled two qubit state with maximally
disordered subsystems. Since $f(\varrho)$ and $\Vert {\mathfrak{A}}(\varrho) \Vert_1$
are both invariant under local unitary operations, we can assume again
that $T(\varrho)$ is diagonal. From Proposition \ref{p2} we know that
$\Vert {\mathfrak{A}}(\varrho) \Vert_1 = \frac{1}{2}(1+ \Vert T(\varrho)
\Vert_1)$. On the other hand an argument similar to the proof of
Equation (\ref{ea1}) in Appendix \ref{a2} leads to
\begin{equation} \label{e8} f(\varrho) = \frac{1}{4} + \max_U \sum_{n=1}^3 \frac{t_{n}}{8}
{\mathrm{tr}}(\sigma_n^{{\mathtt{T}}} U \sigma_n U^\dagger) \end{equation}
where the maximum is over all unitaries $U$ on ${\mathbb{C}}^2$ and
$^{{\mathtt{T}}}$ denotes transposition.
We observe that for
any entangled two qubit state $\varrho$ with maximally disordered
subsystems the number of negative Eigenvalues of
the matrix $T(\varrho)$
is either exactly one or exactly
three. The latter statement
is an immediate consequence of the geometric representation for such
states given in Proposition 3 and Proposition 4 in
Ref.~\onlinecite{Horodecki96}. From Proposition \ref{p3} above and the proof
of Proposition 1 in Ref.~\onlinecite{Horodecki96} (in particular Eq.(13)
there) it follows that there exist a maximally entangled pure state
that compensates the signs of the negative Eigenvalues of $T(\varrho)$.
More precisely, if the signature of $T(\varrho)$ is $(-,-,-)$,
then in Equation \ref{e8} choose $U = e^{i \phi}
\left( \begin{array}{cc}
0 & i \\
-i & 0 \\
\end{array} \right)$. Moreover, for the signatures $(+,+,-)$, $(+,-,+)$ and
$(-,+,+)$ choose $U= e^{i\phi} \left( \begin{array}{cc}
0 & 1 \\
1 & 0 \\
\end{array} \right)$, $U = e^{i \phi} \openone$ and $U =e^{i \phi}\left( \begin{array}{cc}
1 & 0 \\
0 & -1 \\
\end{array} \right)$ respectively.
This shows that $2f(\varrho) = \Vert {\mathfrak{A}}(\varrho)
\Vert_1$. $\Box$

It is worthwhile to note that Proposition \ref{p4} is in
general not true for separable states. To see this consider a
separable state with maximally disordered subsystems for which $T(\varrho)$
has two non-positive Eigenvalues. Such a state
exists by the results of Ref.~\onlinecite{Horodecki96}. To achieve
$2 f(\varrho) = 2 \langle \psi \vert \varrho \vert \psi \rangle =
\Vert {\mathfrak{A}}(\varrho) \Vert_1$ for some maximally
entangled pure state $\vert \psi \rangle$, we need to have, say,
$T(\vert \psi \rangle \langle
\psi \vert) = diag(-1,-1,1)$. However, by the results of
Ref.~\onlinecite{Horodecki96}
there is no state with such a $T$ matrix.

Notice that all the main examples for
two qubit states for which the \texttt{CCN} criterion was
explicitly computed in
Ref.~\onlinecite{Rudolph02} have maximally disordered subsystems. Thus
by Proposition \ref{p2} -- and in accordance with the results of
Ref.~\onlinecite{Rudolph02} -- the \texttt{CCN} criterion is necessary and sufficent
for these states. It is worthwhile to note that there are also families
of two qubit states without maximally disordered subsystems for which
the \texttt{CCN} criterion is a necessary and sufficient condition for separability.
An example is the family of states $\varrho_p = p \vert \psi \rangle
\langle \psi \vert + \frac{1-p}{4} {\openone} \otimes
{\openone}$, where $\vert \psi \rangle$ is a (not necessarily maximally
entangled) pure state and where $p \in \left[ -\frac{1}{3}, 1
\right]$. It is straightforward to check that for this family of
states $\Vert {\mathfrak{A}}(\varrho_p) \Vert_1 \leq 1$ if and only
if $p \leq \frac{1}{4 \sqrt{\alpha_1 \alpha_2}
+1}$ where $(\alpha_1, \alpha_2)$ denote the Schmidt coefficients of $\vert \psi
\rangle$. Invoking the \texttt{PPT}
criterion shows that $\Vert {\mathfrak{A}}(\varrho_p) \Vert_1 \leq
1$ iff $\varrho_p$ is separable.
In view of these examples one may thus conjecture that the \texttt{CCN} criterion
is necessary and sufficient for two qubits. However, it turns out
that this conjecture is not true. A counterexample can easily be
constructed along the lines of
Ref.~\onlinecite{Verstraete01}. Consider a two qubit
state that can be expressed in the form
$\varrho = \frac{1}{4} ({\openone} \otimes
{\openone} + s ({\openone} \otimes \sigma_3) + r (\sigma_3
\otimes {\openone}) + t (\sigma_1 \otimes \sigma_1) - t (\sigma_2
\otimes \sigma_2) + (1+r-s) (\sigma_3 \otimes \sigma_3))$
where $r,s,t \in {\mathbb{R}}$ and where we assume
$s > r$. A straightforward calculation shows that the
partial transpose of this state is positive if and only if
$t=0$. On the other hand $\Vert {\mathfrak{A}}(\varrho) \Vert_1 =
g(s,r) + \vert t \vert$ where
$g(s,r)$ is a non-negative function of $s$ and $r$.
Therefore if we pick appropriate
values for $s$, $r$ and $t$ such that
$g(s,r) < 1$ and such that $0 < \vert t \vert
\leq 1 - g(s,r)$, then the resulting two qubit state is
entangled (as the \texttt{PPT} criterion is necessary and sufficient in
dimension 2 $\times$ 2) but is not detected by the \texttt{CCN} criterion.
A possible choice would be, for instance, $s=\frac{1}{2}, r=\frac{1}{4}$ and
$t= \frac{1}{16}$. Details of
the calculations and the precise form of $g$
can be found in Appendix \ref{a1}. Our example
proves
\begin{prop} \label{p5}
The \texttt{CCN} criterion is not a sufficient criterion for separability
in dimension 2 $\times$ 2.
\end{prop}
\section{\label{s4}The {CCN} criterion in arbitrary dimension}
The aim of the present section is to prove generalized versions
of the Propositions
\ref{p2}, \ref{p3} and \ref{p4} in arbitrary dimensions.
In particular we prove
that $\frac{1}{d}{\mathrm{tr}}({\mathfrak{A}}(\varrho))$ and $\frac{1}{d}
\Vert
{\mathfrak{A}}(\varrho) \Vert_1$ are lower and upper bounds for
the fidelity $f(\varrho)$ respectively.
The examples studied in
Ref.~\onlinecite{Rudolph02} imply that the \texttt{CCN} criterion is not
sufficient for separability in dimension greater
than 2. In this section we
use the
generalized $d$-level spin matrices that were studied in
Ref.~\onlinecite{Bennett93} and \onlinecite{PittengerR00a}.
If we denote the
canonical basis by $(\vert i \rangle)_{i=1}^d$, then the $d$-level
spin matrices are given by
\[ S_{jk} \equiv \sum_{r=0}^{d-1} \exp\left( 2 \pi i jr/d \right) \vert
r \rangle \langle r \oplus k \vert \] where $\oplus$ denotes addition
modulo $d$. It was shown in Ref.~\onlinecite{PittengerR00a} that $\left(
\frac{1}{\sqrt{d}} S_{jk}
\right)_{jk}$ forms an orthonormal basis of the Hilbert-Schmidt
space in $d$ dimensions. Moreover, for $(j,k) \neq (0,0)$ the
matrix $\frac{1}{\sqrt{d}} S_{jk}$ has vanishing trace. We arrange
the matrices $\left( S_{jk} \right)_{(j,k) \neq (0,0)}$ into a $(d^2 -1)$-vector
${\mathbf{S}} = (S_{01}, S_{02}, \cdots, S_{d-1 \, d-1}) =:
(S_1, S_2, \cdots, S_{d^2-1})$. With this notation we can
easily generalize the representation in Equation \ref{e3}. We arrive at that
every bipartite quantum state $\varrho$ on ${\mathbb{C}}^d \otimes
{\mathbb{C}}^d$ can be expanded in Hilbert-Schmidt space as
\begin{widetext}
\begin{equation} \label{e4} \varrho =
\frac{1}{d^2} \left( {\openone} \otimes {\openone}
+ {\mathbf{r}} \cdot \bm{S}
\otimes {\openone} +
{\openone} \otimes {\mathbf{s}} \cdot \bm{S}^* +
\sum_{m,n=1}^{d^2-1} t_{mn} {S}_{n} \otimes {S}^*_m
\right). \end{equation}
\end{widetext}
Here $\mathbf{r}$ and $\mathbf{s}$ are complex vectors in
${\mathbb{C}}^{d^2-1}$ and the $t_{mn}$ form a $(d^2-1) \times (d^2-1)$ complex
matrix $T(\varrho)$. $^*$ denotes complex conjugation.
The reduced states of the subsystems of
$\varrho$ are given by $\varrho_1 = {\mathrm{tr}}_2(\varrho) =
\frac{{\openone} + {\mathbf{r}} \cdot {\mathbf{S}}}{d}$ and
$\varrho_2 = {\mathrm{tr}}_1(\varrho) =
\frac{{\openone} + {\mathbf{s}} \cdot {\mathbf{S}}^*}{d}$.
From Equation \ref{e4} we infer \begin{widetext}
\begin{equation} \label{e5} {\mathfrak{A}}(\varrho) =
\frac{1}{d^2} \left( \vert {\openone} \rangle \langle {\openone}
\vert
+ \sum_i {{r_i}} \vert {S_i} \rangle \langle
{\openone} \vert + \sum_i {{s_i}}
\vert {\openone} \rangle \langle {S_i} \vert +
\sum_{m,n=1}^{d^2-1} t_{mn} \vert {S}_{n} \rangle \langle {S}_m
\vert
\right). \end{equation} \end{widetext}
We now
wish to relate the operator ${\mathfrak{A}}(\varrho)$ to the
fidelity $f(\varrho)$ of $\varrho$. The results of
Ref.~\onlinecite{Rudolph02,Horodecki97} imply that if $\varrho$ is pure
or an isotropic state, then $df(\varrho) = \Vert {\mathfrak{A}}(\varrho)
\Vert_1$. In general we will see that equality does not hold.
However, Proposition \ref{p3} immediately generalizes to arbitrary
dimension, i.e., we have
\begin{prop} \label{p6} Let $\varrho \in {\mathtt{T}}({\mathbb{C}}^d \otimes
{\mathbb{C}}^d)$ be a bipartite state. Then
$d f(\varrho) \leq \Vert {\mathfrak{A}}(\varrho) \Vert_1$. \end{prop}
For a proof we refer to Appendix \ref{a2}. Moreover we
have the following proposition
\begin{prop} \label{p8}
Let $\varrho$ be a bipartite state on ${\mathbb{C}}^d \otimes
{\mathbb{C}}^d$. Then \[ d f(\varrho) = \max_U \vert
{\mathrm{tr}}
{\mathfrak{A}}((\openone \otimes U)\varrho
(\openone \otimes U^\dagger))
\vert \] where the maximum is over all unitary operators $U$ on
${\mathbb{C}}^d$. Moreover,
$d \langle \Psi_+ \vert \varrho \vert \Psi_+ \rangle =
{\mathrm{tr}}({\mathfrak{A}}(\varrho))
\leq d f(\varrho).$ If in addition ${\mathfrak{A}}(\varrho) \geq
0$,
then
$d f(\varrho) = \vert {\mathrm{tr}}({\mathfrak{A}}(\varrho))
\vert =
\Vert {\mathfrak{A}}(\varrho) \Vert_1$.
\end{prop}
\emph{Proof}:
From Equation \ref{e5} we infer
${\mathrm{tr}}({\mathfrak{A}}(\varrho)) = \frac{1}{d}
\left( 1+ \sum_{n} t_{nn} \right).$ On the other hand note that
\cite{DonaldHR02}
$f(\varrho) = \max_\psi \langle \psi \vert \varrho \vert \psi \rangle =
\max_U \langle \Psi_+ \vert (\openone \otimes U) \varrho (\openone
\otimes U^\dagger) \vert \Psi_+ \rangle$. The first maximum is
with respect to all maximally entangled states $\psi$ while the
second is with respect to all unitary operators on
${\mathbb{C}}^d$. Moreover, $\vert \Psi_+ \rangle
= \frac{1}{\sqrt{d}} \sum_i \vert i, i \rangle$ where $(\vert i \rangle)$
denotes the canonical real basis of ${\mathbb{C}}^d$.
A straightforward calculation shows that
\begin{equation}
f(\varrho) = \frac{1}{d^2}\left(1 + \max_U \frac{1}{d}
\sum_{mn} t_{mn}
\langle S_m \vert U^{\mathtt{T}} S_n U^* \rangle
\right).
\end{equation}
This implies the variational expression in Proposition
\ref{p8} and also that ${\mathrm{tr}}({\mathfrak{A}}(\varrho)) \leq
df(\varrho)$. Moreover we find ${\mathrm{tr}}({\mathfrak{A}}(\varrho)) =
d \langle \Psi_+ \vert \varrho \vert \Psi_+ \rangle$ (corresponding to
$U= \openone$).
If ${\mathfrak{A}}(\varrho) \geq 0$, then
${\mathrm{tr}}({\mathfrak{A}}(\varrho))
\leq d f(\varrho)
 \leq \Vert {\mathfrak{A}}(\varrho) \Vert_1 =
 {\mathrm{tr}}({\mathfrak{A}}(\varrho)).$ $\Box$
\begin{co}
Let $\varrho$ be a bipartite state on ${\mathbb{C}}^d \otimes
{\mathbb{C}}^d$. If ${\mathrm{tr}}({\mathfrak{A}}(\varrho)) >1$,
then $\varrho$ is distillable.
\end{co}
\emph{Proof}: This follows immediately from Proposition \ref{p8}
and the results of Ref.~\onlinecite{Horodecki97}.

\begin{co}
Let $\varrho$ be a bipartite state on ${\mathbb{C}}^d \otimes
{\mathbb{C}}^d$. Then ${\mathrm{tr}}({\mathfrak{A}}(\varrho)) \geq
0$.
\end{co} Note that ${\mathfrak{A}}(\varrho)$ is in general
not Hermitean.
 The
following proposition is our generalization of Proposition
\ref{p2}.
\begin{prop} \label{p7}
Let $\varrho$ a bipartite state with maximally disordered
subsystems. Then $\Vert {\mathfrak{A}}(\varrho) \Vert_1 = \frac{1}{d}
\left( 1+ \Vert T(\varrho) \Vert_1 \right)$.
If $T(\varrho) \geq 0 $
and $\Vert {\mathfrak{A}}(\varrho) \Vert_1 >1$, then
$\varrho$ is distillable.
\end{prop}
\emph{Proof}: Let $\varrho$ be a bipartite
state with maximally disordered
subsystems. Then as in the proof of Proposition \ref{p2}
\[ {\mathfrak{A}}(\varrho) = \frac{1}{d^2}\left( \vert \openone
\rangle \langle \openone \vert + \sum_{m,n=1}^{d^2-1} t_{mn} \vert
S_n \rangle \langle S_m \vert \right). \]
Since $(\frac{1}{\sqrt{d}} \vert \openone \rangle, \frac{1}{\sqrt{d}}
\vert S_n \rangle)$ forms an orthonormal basis of the
Hilbert-Schmidt space in dimension $d$, we find
that $\Vert
{\mathfrak{A}}(\varrho) \Vert_1 = \frac{1}{d} (1+\Vert T(\varrho)
\Vert_1)$. This proves the first half of Proposition \ref{p7}.
From Equation \ref{e5} we see that for
states $\varrho$ with maximally disordered subsystems
$T(\varrho) \geq
0$ if and only if ${\mathfrak{A}}(\varrho) \geq 0$.
Now if $T(\varrho) \geq 0$, then by Proposition
\ref{p8} $df(\varrho) = \Vert {\mathfrak{A}}(\varrho) \Vert_1$.
Thus $\Vert {\mathfrak{A}}(\varrho) \Vert_1 > 1$ is equivalent to
$f(\varrho) > \frac{1}{d}.$ By the results of
Ref.~\onlinecite{Horodecki97} this implies that $\varrho$ is
distillable. This proves the proposition. $\Box$

\section{\label{s5}The {CCN} criterion under local operations}
In the paradigmatic situation studied in quantum information
theory two parties, traditionally called Alice and Bob, share
parts of composite quantum systems and are able to
perform local operations on their respective parts and communicate
classically. An essential requirement for measures of entanglement
is to be non-increasing under \texttt{LQCC} operations, i.e.,
operations that can be implemented locally with classical
communication between the parties. In the present
section we study the behaviour of the quantity $\Vert {\mathfrak{A}}(\varrho)
\Vert_1$ under local operations.
An operation is a completely positive
linear map $\Lambda$ that is trace non-increasing for positive
operators. In the following we are only interested in trace
preserving operations.
Such quantum operations are all those operations
that can be composed out of the following elementary operations
\cite{DonaldHR02}:
(O1) adding an uncorrelated ancilla system;
(O2) tracing out part of the system;
(O3) unitary transformations;
(O4) L\"uders-von Neumann measurements:
$\Lambda_{{\mathrm{LvN}}} : {\mathtt{T}}({\mathtt{H}}) \to
{\mathtt{T}}({\mathtt{H}}), \Lambda_{{\mathrm{LvN}}}(\rho)
= \sum_{i=1}^r P_i \rho
P_i$ where $(P_i)_{i=1}^r$ is a complete
sequence of pairwise
orthogonal projection operators on ${\mathtt{H}}$.
\begin{prop} \label{p9}
The quantity $\Vert {\mathfrak{A}}(\varrho) \Vert_1$ remains
invariant under local operations of the type (O3). It is
non-increasing under local operations of type (O1) and
(O4). $\Vert {\mathfrak{A}}(\varrho)
\Vert_1$ may increase, decrease or stay invariant under local operations
of type (O2). \end{prop}

\begin{co} \label{p10}
The \texttt{CCN} criterion is not invariant under local
operations.
\end{co}
The statement of the Corollary means
that if $\varrho$ is a state satisfying, say, $\Vert
{\mathfrak{A}}(\varrho) \Vert_1 \leq 1$, then there may be a state
$\Lambda(\varrho)$ obtained from $\varrho$ by a local trace non-increasing
operation
$\Lambda$ such that $\Vert {\mathfrak{A}}(\Lambda(\varrho))
\Vert_1 > 1.$

\noindent \emph{Proof of Proposition \ref{p9}}:
The invariance of $\Vert {\mathfrak{A}}(\varrho)
\Vert_1$ under local unitary operations is an immediate
consequence of the representation in Equation \ref{tau}.
Similarly, it is immediate from Equation \ref{tau} that $\Vert {\mathfrak{A}}(\varrho)
\Vert_1$ is non-increasing under adding a local ancilla (O1). To
see that $\Vert {\mathfrak{A}}(\varrho) \Vert_1$ is non-increasing
under operations of type (O4), let $(P_k)_k$ be a complete family
of mutually orthogonal projectors on ${\mathbb{C}}^d$ and let
$\Lambda_{\text{LvN}}(\varrho) \equiv \sum_k (P_k \otimes \openone)
\varrho (P_k \otimes \openone)$. Then using Equation \ref{tau}
yields \begin{widetext}
\begin{eqnarray*}
\Vert {\mathfrak{A}}(\Lambda_{\text{LvN}}(\varrho)) \Vert_1 & \leq
& \inf \left\{ \sum_i \Vert \sum_k P_k x_i P_k
\Vert_2 \Vert y_i
\Vert_2 : \varrho = \sum_i x_i \otimes y_i \right\} \\
& \leq & \inf \left\{ \sum_i \Vert x_i \Vert_2 \Vert y_i
\Vert_2 : \varrho = \sum_i x_i \otimes y_i \right\}
= \Vert {\mathfrak{A}}(\varrho) \Vert_1
\end{eqnarray*} \end{widetext}
where in the second line we used that the
Hilbert-Schmidt norm is non-increasing under \emph{pinching},
i.e., $\Vert \sum_k P_k \sigma P_k \Vert_2 \leq \Vert \sigma
\Vert_2$ for all families of mutually orthogonal projectors with $\sum_k P_k
=1$ and all $\sigma$.
Finally consider two bipartite states $\varrho_1$ and $\varrho_2$
that satisfy $\Vert {\mathfrak{A}}(\varrho_1) \Vert_1 < 1$ and $\Vert
{\mathfrak{A}}(\varrho_2) \Vert_1 > 1$. Then
$\Vert {\mathfrak{A}}(\varrho_1 \otimes \varrho_2) \Vert_1 =
\Vert {\mathfrak{A}}(\varrho_1) \Vert_1 \Vert
{\mathfrak{A}}(\varrho_2) \Vert_1$. It is immediate that if Alice
and Bob locally trace out $\varrho_1$, then the value of
$\Vert {\mathfrak{A}}(\varrho) \Vert_1$ will increase, while
tracing out $\varrho_2$ decreases $\Vert {\mathfrak{A}}(\varrho)
\Vert_1$. [If $\varrho_1$ would satisfy $\Vert {\mathfrak{A}}(\varrho_1)
\Vert_1 = 1$, tracing out $\varrho_1$ would obviously
leave the value of
$\Vert {\mathfrak{A}}(\varrho)
\Vert_1$ invariant.] $\Box$ \\

\noindent \emph{Proof of Corollary \ref{p10}}:
The argument in the proof of Proposition \ref{p9}
also implies that the \texttt{CCN} criterion is not invariant
under local operations. To see this, choose $\varrho_1$ and
$\varrho_2$ such that
$\Vert {\mathfrak{A}}(\varrho_1 \otimes \varrho_2) \Vert_1 =
\Vert {\mathfrak{A}}(\varrho_1) \Vert_1 \Vert
{\mathfrak{A}}(\varrho_2) \Vert_1 < 1$ and
$\Vert {\mathfrak{A}}(\varrho_2) \Vert_1 >1$. I.e., the state
$\varrho_1 \otimes
\varrho_2$ satisfies the \texttt{CCN} criterion. Tracing out $\varrho_1$
leaves Alice and Bob with $\varrho_2$, i.e., with a state that
violates the \texttt{CCN} criterion. $\Box$

Proposition \ref{p9} and Corollary \ref{p10}
show that an entangled state $\varrho$ that
satisfies the \texttt{CCN} criterion may be transformed into a state
violating it by locally tracing out part of the system. This
suggests the following extension of the \texttt{CCN} criterion.
\begin{cri} \label{c2} Consider the quantity
\[ {\mathbb{A}}(\varrho) := \sup_{K_A,K_B} \Vert
{\mathfrak{A}}({\mathrm{tr}}_{K_a \otimes K_B}(\varrho)) \Vert_1
\] where the supremum is over all local spaces $K_A$ and $K_B$
(on Alice's and Bob's side respectively) that can be traced out
locally.
The extended \texttt{CCN} criterion
asserts that if $\varrho$ is separable, then ${\mathbb{A}}(\varrho) \leq 1$.
Whenever a quantum state $\varrho$ satisfies
${\mathbb{A}}(\varrho) > 1$, this signals that $\varrho$
is entangled.
\end{cri}
This new criterion is stronger than the \texttt{CCN} criterion. A trivial
example has been given above in the proof of Corollary \ref{p10}.
Since there are infinitely many ways of realizing an isomorphism
${\mathbb{C}}^{d_1d_2} \simeq {\mathbb{C}}^{d_1} \otimes
{\mathbb{C}}^{d_2}$ the quantity ${\mathbb{A}}(\varrho)$ will in general
not be computable and thus the criterion \ref{c2} is not fully
operational. By fixing an isomorphism it is obviously always
possible to pass to a
weaker but operational criterion.
However, we
have not yet identified
a non trivial example where the extended
criterion detects entanglement that is not already
detected by the \texttt{CCN}
criterion. This problem is thus left as an open problem.
\acknowledgments
The author would like to thank Shashank Virmani for stimulating
discussions.
Funding by the EC project ATESIT (contract
IST-2000-29681) is gratefully acknowledged.
\appendix
\section{\label{a2}Proof of Propositions \ref{p3} and \ref{p6}}
First we extend the definition of fidelity to arbitrary trace
class operators on $\varpi \in {\mathtt{T}}({\mathbb{C}}^d \otimes
{\mathbb{C}}^d)$ by \[ f(\varpi) := \max_{\Psi} \vert \langle \Psi
\vert \varpi \vert \Psi \rangle \vert \] where the maximum is over
all maximally entangled pure states $\vert \Psi \rangle$.
Every maximally entangled wavefunction is of the form
$\vert \Psi \rangle = ({\openone} \otimes U) \vert \Psi_+ \rangle$ for some
unitary $U$ \cite{DonaldHR02}.
It is
straightforward to check that for all operators of the form
$\varpi_1 \otimes
\varpi_2$ we have \begin{equation} \label{ea1} f(\varpi_1 \otimes
\varpi_2) = \frac{1}{d}
\max_{U} \vert
{\mathrm{tr}}(\varpi_1^{\mathtt{T}} U \varpi_2 U^\dagger) \vert
\end{equation} where $^{{\mathtt{T}}}$ denotes transposition.
This implies that
\[ f(\varpi_1 \otimes \varpi_2) \leq \frac{1}{d} \Vert
\varpi_1
\Vert_2 \Vert \varpi_2 \Vert_2. \] In other words $d f$ satisfies
the \emph{subcross property} with respect to the Hilbert-Schmidt
norm $\Vert \cdot \Vert_2$. This implies immediately that
$d f(\varpi) \leq \Vert {\mathfrak{A}}(\varpi) \Vert_1$, as $\Vert
{\mathfrak{A}}(\varpi) \Vert_1$ is the greatest cross norm with
respect to the Hilbert-Schmidt norm. (Note that this should be carefully distinguished
from the greatest cross norm with respect to the trace class norm that was
studied in Ref.~\onlinecite{Rudolph00}). Namely let $\varpi =
\sum_{i=1}^k x_i \otimes y_i$ be a decomposition of $\varpi$ into
a finite sum of simple tensors, then
\[ f(\varpi) \leq \sum_{i=1}^k f(x_i \otimes y_i) \leq \frac{1}{d}
\sum_{i=1}^k \Vert x_i \Vert_2 \Vert y_i \Vert_2. \] Taking the
infimum over all possible finite decompositions on the right hand
side yields (compare Equation \ref{tau})
\[ d f(\varpi) \leq \Vert {\mathfrak{A}}(\varpi) \Vert_1. \]
\section{\label{a1}The counterexample in dimension $2 \times 2$}
The matrix representation of the state $\varrho = \frac{1}{4} ({\openone} \otimes
{\openone} + s ({\openone} \otimes \sigma_3) + r (\sigma_3
\otimes {\openone}) + t (\sigma_1 \otimes \sigma_1) - t (\sigma_2
\otimes \sigma_2) + (1+r-s) (\sigma_3 \otimes \sigma_3))$ in the canonical
basis is given
by \[ \varrho = \frac{1}{2} \left( \begin{array}{cccc}
1+r & 0 & 0 & t \\
0 & 0 & 0 & 0 \\
0 & 0 & s-r & 0 \\
t & 0 & 0 & 1-s \\
\end{array} \right). \]
The Eigenvalues are given by $\lambda_1 = 0, \lambda_2 = \frac{s-r}{2}, \lambda_{3,4} =
\frac{1}{2} + \frac{r-s}{4} \pm \frac{1}{2} \sqrt{t^2 + \frac{(s+r)^2}{4}}$. $\varrho$ is
a state if the parameters $s,r,t$ are chosen such that each
$\lambda_i \geq
0$. We assume that $s > r$. By considering the subsystems of $\varrho$, we see that
$\vert s \vert \leq 1$ and $\vert r \vert \leq 1$.
The Eigenvalues of the partial transpose of $\varrho$ are
easily confirmed to be $\lambda_1 = \frac{1 +r}{2}, \lambda_2= \frac{1-s}{2},
\lambda_{3,4}
= \frac{s-r}{4} \pm \frac{1}{2}
\sqrt{\frac{(s-r)^2}{4} + t^2}$. Therefore
$\varrho^{{\mathtt{T}}_2}$ has a
negative Eigenvalue if and only if $t \neq
0$.
Now ${\mathfrak{A}}(\varrho)$ is given by \begin{widetext}
\[
{\mathfrak{A}}(\varrho) =
\frac{1}{4} \left( \left\vert
{\openone} \right\rangle \left\langle
{\openone} \right\vert +
s \left\vert {\sigma_3} \right\rangle \left\langle
{\openone} \right\vert +
r \left\vert
{\openone} \right\rangle \left\langle
{\sigma_3} \right\vert +
(1+r-s) \left\vert {\sigma_3} \right\rangle
\left\langle
{\sigma_3} \right\vert +
t \left\vert {\sigma_1} \right\rangle
\left\langle
{\sigma_1} \right\vert +t
\left\vert {\sigma_2} \right\rangle
\left\langle
{\sigma_2} \right\vert
\right). \] \end{widetext} The matrix representation of ${\mathfrak{A}}(\varrho)$
is
 \[ {\mathfrak{A}}(\varrho) = \frac{1}{2} \left( \begin{array}{cccc}
1 & 0 & 0 & r \\
0 & t & 0 & 0 \\
0 & 0 & t & 0 \\
s & 0 & 0 & 1+r-s \\
\end{array} \right) \]
The trace class norm of this operator is easily computed.
We set
$\psi(s,r) \equiv (1+r)^2 + (s-r)^2 + (1-s)^2$. The
Eigenvalues of the operator ${\mathfrak{A}}(\varrho)^\dagger
{\mathfrak{A}}(\varrho)$ are then
\begin{eqnarray*} \lambda_1 & = &
\frac{1}{8} \left(\psi(s,r) +
\sqrt{\psi(s,r)^2 - 4 (1+r)^2(1-s)^2} \right) \\
\lambda_4 & = &
\frac{1}{8} \left(\psi(s,r) -
\sqrt{\psi(s,r)^2 - 4 (1+r)^2(1-s)^2} \right)
\end{eqnarray*} and $\lambda_2 = \lambda_3
= \frac{t^2}{4}$.
Therefore if we set $g(s,r) := \sqrt{\lambda_1} +
\sqrt{\lambda_4}$, we arrive at
\[ \Vert {\mathfrak{A}}(\varrho) \Vert_1 = g(s,r) + \vert t \vert. \]

\end{document}